\documentstyle[aps,preprint,fleqn,epsfig]{revtex}
\tightenlines
\begin{document}
\title{Studies of the Electric Dipole Transition of Deformed Rare-earth
nuclei}
\author{
H.Y.Ji$^1$ ,
G.L.Long$^{1,2,3}$\thanks{Corresponding author: Gui Lu
Long, Department of Physics, Tsinghua University, Beijing, 100084, P. R.
China. Email: gllong@mail.tsinghua.edu.cn},
E. G. Zhao$^{1,2}$ and S. W. Xu$^4$ }
\address{
$^{1}$Department of Physics, Tsinghua University, Beijing 100084, China\\
$^2$Institute of Theoretical Physics, Chinese Academy of Sciences,
Beijing, China\\
$^3$Center of Theoretical Nuclear Physics, National Laboratory\\
 of Heavy Ion Accelerator, Chinese Academy of Sciences, Lanzhou,
 730000,P R China\\
$^4$ Institute of Modern Physics, Chinese Academy of Sciences, Lanzhou,
730000, P. R. China}
\date{\today}
\maketitle

\begin{abstract}
Spectrum and electric dipole transition rates and relative
intensities in $^{152-154}$Sm, $^{156-160}$Gd, $^{160-162}$Dy
are studied in the framework of the interacting boson model 
with s,p,d,f bosons. It is found that E1 transition data 
among the
low-lying levels are in good agreement with the SU(3) dynamical
symmetry of the spdf interacting boson model proposed by Engel and
Iachello
to
describe collective rotation with octupole vibration.
These results  show that these 
nuclei have SU(3) dynamic symmetry to a good approximation. 
Also in this work many algebraic expressions for
electric dipole transitions  in the SU(3) limit of the spdf-IBM
have been obtained. 
These formulae together with the formulae given previously  
exhaust nearly all the E1 transitions for low-lying negative 
parity states. They are useful in analyzing experimental data.
\end{abstract}


\section{Introduction}

There have been intensive interests in the studies of octupole degree of
 freedom in nuclear structure 
recently both experimentally
\cite{r1,r2,r3,r4,r5,r6,r7,r20,r21}
and theoretically
\cite{r8,r9,r10,r11,r11a,kus,r12,r13,r14,r15,r16,r17,r18,r19,r19a,r19b,r19c,r22}. 

In the boson model, negative parity
states are described by the 
spdf interacting boson model\cite{r8,r9,r10,r11a,r11,r12,r13,r15,r16,r17,r23} 
or the sdf-IBM\cite{r18,r19,r19a,r19b,r19c}. Otsuka \cite{r11a} showed
microscopically that $p$ and $f$ bosons are important in the low-lying
negative parity states of even-even nuclei.
The spdf IBM has been
successful in the description of negative parity states 
in the Ba
and rare-earth region\cite{r13,r15,r16}. 
The advantages of the algebraic approach is that it offers
 analytical expressions for the energy levels, 
electromagnetic transitions
 and many other quantities.  
The prediction and later experimental confirmation of the O(6) 
limit has been a well-known example of the 
success of the IBM\cite{r24}.

 The spdf IBM SU(3) limit is a dynamical symmetry describing
 octupole vibration in deformed nuclei proposed by Engel and 
Iachello\cite{r10,r11}. 
It has been pointed that $^{232}$U or other actinide nuclei may be 
 candidates for the SU(3) limit\cite{r10,r11,r12}. 
While the spectrum agrees with the
theoretical calculation very well, there are very few electromagnetic 
transition data. Thus it is difficult to check on the validity
 of the dynamic symmetry properties, in particular on the electric dipole
transitions connecting positive and negative parity states.
This is because finding nucleus
 with both good rotational feature in spectrum and with ample data of 
electric dipole transition is difficult.
Besides, in real nucleus dynamical symmetry is 
usually somehow broken, for instance, even one of the 
best candidate of the sdg IBM SU(3) dynamic symmetry has some degree of 
symmetry-breaking\cite{r25}.
But dynamic symmetries are very useful even though they are broken.
They can be used to 
classify states and characterize the collective features of a nucleus. 
To a given nucleus near a dynamical symmetry, 
at first the gross structure is dictated by the dynamical symmetry. 
Then detailed structure of the
nucleus can be attributed to symmetry breaking,
and its description 
is the task of a more elaborate study.
In addition, the dynamical symmetries can be used as landmarks in the
nuclear chart table to classify typical collective motions, 
and other vast majority of nuclei can be put into categories of 
transition between two dynamical symmetries. Such a scheme is very
successful in the descriptions of the positive low-lying states of
even-even nuclei\cite{r25a}. 
In this work we have studied the negative parity states of
the rare-earth nuclei, namely 
$^{152-154}$Sm, $^{156-160}$Gd, $^{160-162}$Dy
using the
SU(3) limit of the spdf interacting boson model.  These nuclei are
deformed and have ample data of electric dipole transition. 
We have found in this work 
that to a good approximation, the structure of these
rare-earth nuclei can be well described by the SU(3) limit of the spdf
interacting boson model. 
The paper is organized in the following. 
In section \ref{s2}, we give a brief
description of the model and  the necessary expressions of
electric dipole transitions.
In section \ref{s3}, we apply the results to $^{152}$Sm on
the spectrum and E1 transitions.
In section \ref{s4}, we study other deformed rare-earth nuclei.
Finally
 we give a discussion and a summary in section \ref{s5}. 

\section{The model}
\label{s2}

\subsection{The energy eigenvalues}
\label{s2.1}
 The energy eigenvalues have been discussed in Ref.\cite{r10,r13}. 
In the SU(3) 
limit, the group chain can be written as
\begin{eqnarray}
&&U(16)\supset U(6)\otimes U(10)\supset SU_{sd}(3)\otimes SU_{pf}(3)
\supset SU_{spdf}(3)\supset O(3)\nonumber\\
&& \;\;\;n\;\;\;\;\;\;\;\;\;\;n_+\;\;\;\;\;\;\;n_-\;\;\;\;\;\;(\lambda_+
\mu_+)\;\;\;\;\;\;(\lambda_- \mu_-)\;\;\;\;\;\;\;\;(\lambda \mu)\;\;\;\;
\;\;\;\;\;\;\;L,
\end{eqnarray}
the energy eigenvalue is
\begin{eqnarray}
E=\epsilon_-N_{pf}+a_1C_{2SU_+}+a_2C_{2SU_-}+a_3C_{2SU(3)}+a_4L(L+1).
\end{eqnarray}

 The g.s.-band,$\beta$-band and $\gamma$-band are generated from
$(2n,0),(2n-4,2)K=0$ and $(2n-4,2)K=2$ respectively, with $n$ sd-bosons. The 
low-lying negative parity are generalized by the SU(3) IR from the 
decomposition of (2n,0)$\otimes$(3,0); that is $K^p=0^-$ from (2n+3,0)
 and $K^p=1^-$ from (2n+1,1) respectively. For the low-lying states, 
we are interested in only the g.s., $\beta$, $\gamma$,$0^-$ and $1^-$ bands.

The wave function are given by 
\begin{eqnarray}
&&(1)ground\; state\; band\; \vert(2n,0)_+LM\rangle,\nonumber\\
&&(2)\beta\;band\;\vert(2n-4,2)_+K=0LM\rangle,\nonumber\\
&&(3)\gamma\;band\;\vert(2n-4,2)_+K=2LM\rangle,\nonumber\\
&&(4)K^p=0^--band\;\vert(2n-2,0)_+(3,0)_-(2n+1,0)LM\rangle,\nonumber\\
&&(5)K^p=1^--band\;\vert(2n-2,0)_+(3,0)_-(2n-1,1)LM\rangle.
\end{eqnarray}

The value of $a_2$ is taken zero, since its effect 
in the spectrum is the same as that of $\epsilon_-$ term for 
the low-lying states with only one pf-boson. The parameters are 
then determined by experimental data.

\subsection{The E1 transition matrix elements}
\label{s2s2}

 Some of the E1 transition formulae have been given in Ref.\cite{r23}. 
With the
method described there, we calculate the following formulae for the
E1 transition, which are needed for a comparison with experiment, and the
required SU(3) Wigner coefficients can be found in Ref.\cite{r27,r28}.

\noindent (a)(2n+3,0)K=0 $L^- \rightarrow$ (2n-2,2)K=0 $(L-1)^+$
 transitions, e.g. $1^-\rightarrow0^+$,

\begin{eqnarray}
\lefteqn{\langle(L-1)^{+}\vert\vert(s^{\dagger}\tilde{p}+p^{\dagger}
\tilde{s})^{1}
\vert\vert{L^-}\rangle={\frac{1}{(2n+1)}}\times} &&\nonumber \\
&&\sqrt{\frac{(2n-L+3)(2n+L+2)(2n+L+4)L\;
 \varphi(2n-1,L-1)}{20n(2n+3)}},
\end{eqnarray}
where
\begin{eqnarray}
\varphi(\lambda,L)=2(\lambda+1)^2-L(L+1),
\end{eqnarray}
\begin{eqnarray}
\lefteqn{\langle(L-1)^{+}\vert\vert(d^{\dagger}\tilde{p}+p^{\dagger}
\tilde{d})^{1}
\vert\vert{L^-}\rangle={\frac{8n^2-L(L-1)}{5(2n+1)}}\times}&&\nonumber \\
&&\sqrt{\frac{(2n-L+3)(2n+L+2)(2n+L+4)L
}{4n(2n+3)\;\varphi(2n-1,L-1)}},
\end{eqnarray}
\begin{eqnarray}
\lefteqn{\langle(L-1)^{+}\vert\vert(d^{\dagger}\tilde{f}+f^{\dagger}
\tilde{d})^{1}
\vert\vert{L^-}\rangle={-\frac{8n^2-L(L-1)}{5(2n+1)}}\times}&&\nonumber \\
&&\sqrt{\frac{3(2n-L+3)(2n+L+2)(2n+L+4)L
}{28n(2n+3)\;\varphi(2n-1,L-1)}}.
\end{eqnarray}

\noindent (b)(2n+3,0)K=0 $L^- \rightarrow$ (2n-2,2)K=0 $(L+1)^+$
transitions,
 e.g.
 $1^-\rightarrow2^+$,

\begin{eqnarray}
\lefteqn{\langle(L+1)^{+}\vert\vert(s^{\dagger}\tilde{p}+
p^{\dagger}\tilde{s})^{1}
\vert\vert{L^-}\rangle={-\frac{1}{(2n+1)}}\times}&&\nonumber \\
&&\sqrt{\frac{(2n-L+1)(2n-L+3)(2n+L+4)(L+1)\;\varphi(2n-1,L+1)}{20n(2n+3)}},
\end{eqnarray}
\begin{eqnarray}
\lefteqn{\langle(L+1)^{+}\vert\vert(d^{\dagger}\tilde{p}+
p^{\dagger}\tilde{d})^{1}
\vert\vert{L^-}\rangle={-\frac{8n^2-(L+1)(L+2)}{5(2n+1)}}\times}&&\nonumber \\
&&\sqrt{\frac{(2n-L+1)(2n-L+3)(2n+L+4)(L+1)}{4n(2n+3)\;\varphi(2n-1,L+1)}},
\end{eqnarray}
\begin{eqnarray}
\lefteqn{\langle(L+1)^{+}\vert\vert(d^{\dagger}\tilde{f}+
f^{\dagger}\tilde{d})^{1}
\vert\vert{L^-}\rangle={\frac{8n^2-(L+1)(L+2)}{5(2n+1)}}\times}&&\nonumber \\
&&\sqrt{\frac{3(2n-L+1)(2n-L+3)(2n+L+4)(L+1)}{28n(2n+3)\;\varphi(2n-1,L+1)}}.
\end{eqnarray}

\noindent (c)(2n+3,0)K=0 $L^- \rightarrow$ (2n-2,2)K=2 $(L-1)^+$
 transitions, e.g. $3^-\rightarrow2^+$,

\begin{eqnarray}
\lefteqn{\langle(L-1)^{+}\vert\vert(s^{\dagger}\tilde{p}+
p^{\dagger}\tilde{s})^{1}\vert\vert{L^-}\rangle=0,}
\end{eqnarray}
\begin{eqnarray}
\lefteqn{\langle(L-1)^{+}\vert\vert(d^{\dagger}\tilde{p}+
p^{\dagger}\tilde{d})^{1}\vert\vert{L^-}\rangle=0,}
\end{eqnarray}
\begin{eqnarray}
\lefteqn{\langle(L-1)^{+}\vert\vert(d^{\dagger}\tilde{f}+
f^{\dagger}\tilde{d})^{1}\vert\vert{L^-}\rangle=0.}
\end{eqnarray}

\noindent (d)(2n+3,0)K=0 $L^- \rightarrow$ (2n-2,2)K=2 $L^+$
 transitions, e.g. $3^-\rightarrow3^+$,

\begin{eqnarray}
\lefteqn{\langle L^{+}\vert\vert(s^{\dagger}\tilde{p}+
p^{\dagger}\tilde{s})^{1}\vert\vert{L^-}\rangle=0,}
\end{eqnarray}
\begin{eqnarray}
\lefteqn{\langle L^{+}\vert\vert(d^{\dagger}\tilde{p}+
p^{\dagger}\tilde{d})^{1}\vert\vert{L^-}\rangle=0,}
\end{eqnarray}
\begin{eqnarray}
\lefteqn{\langle L^{+}\vert\vert(d^{\dagger}\tilde{f}+
f^{\dagger}\tilde{d})^{1}\vert\vert{L^-}\rangle=0.}
\end{eqnarray}

\noindent (e)(2n+3,0)K=0 $L^- \rightarrow$ (2n-2,2)K=2 $(L+1)^+$
 transitions, e.g. $1^-\rightarrow2^+$,

\begin{eqnarray}
\lefteqn{\langle(L+1)^{+}\vert\vert(s^{\dagger}\tilde{p}+
p^{\dagger}\tilde{s})^{1}\vert\vert{L^-}\rangle=0,}
\end{eqnarray}
\begin{eqnarray}
\lefteqn{\langle(L+1)^{+}\vert\vert(d^{\dagger}\tilde{p}+
p^{\dagger}\tilde{d})^{1}\vert\vert{L^-}\rangle=0,}
\end{eqnarray}
\begin{eqnarray}
\lefteqn{\langle(L+1)^{+}\vert\vert(d^{\dagger}\tilde{f}+
f^{\dagger}\tilde{d})^{1}\vert\vert{L^-}\rangle=0.}
\end{eqnarray}

\noindent (f)(2n+1,1)K=1 $L^- \rightarrow$ (2n+2,0)K=0 $(L-1)^+$
 transitions, e.g. $1^-\rightarrow0^+$,

\begin{eqnarray}
\lefteqn{\langle(L-1)^{+}\vert\vert(s^{\dagger}\tilde{p}+
p^{\dagger}\tilde{s})^{1}
\vert\vert{L^-}\rangle={-\frac{(2n-3L+3)(2n+L+2)}{(2n+1)}}\times}&&
\nonumber \\
&&\sqrt{\frac{(2n-L+3)(L+1)}{60n(2n+3)}},
\end{eqnarray}
\begin{eqnarray}
\lefteqn{\langle(L-1)^{+}\vert\vert(d^{\dagger}\tilde{p}+
p^{\dagger}\tilde{d})^{1}
\vert\vert{L^-}\rangle={-\frac{4n^2+2(4L+5)n-3(L-1)(L+2)}{5(2n+1)}}
\times}&&\nonumber \\
&&\sqrt{\frac{(2n-L+3)(L+1)}{12n(2n+3)}},
\end{eqnarray}
\begin{eqnarray}
\lefteqn{\langle(L-1)^{+}\vert\vert(d^{\dagger}\tilde{f}+
f^{\dagger}\tilde{d})^{1}
\vert\vert{L^-}\rangle={-\frac{16n^2+12Ln+3(L-1)(L+2)}{5(2n+1)}}\times}
&&\nonumber \\
&&\sqrt{\frac{(2n-L+3)(L+1)}{28n(2n+3)}}.
\end{eqnarray}

\noindent (g)(2n+1,1)K=1 $L^- \rightarrow$ (2n+2,0)K=0 $L^+$
 transitions, e.g. $2^-\rightarrow2^+$,

\begin{eqnarray}
\lefteqn{\langle L^{+}\vert\vert(s^{\dagger}\tilde{p}+
p^{\dagger}\tilde{s})^{1}\vert\vert{L^-}\rangle=-\frac{(2n-L+2)(2n+L+3)}
{(2n+1)}\sqrt{\frac{2L+1}{60n}},}
\end{eqnarray}
\begin{eqnarray}
\lefteqn{\langle L^{+}\vert\vert(d^{\dagger}\tilde{p}+
p^{\dagger}\tilde{d})^{1}\vert\vert{L^-}\rangle=-\frac{(2n-L+2)(2n+L+3)}
{5(2n+1)}\sqrt{\frac{2L+1}{12n}},}
\end{eqnarray}
\begin{eqnarray}
\lefteqn{\langle L^{+}\vert\vert(d^{\dagger}\tilde{f}+
f^{\dagger}\tilde{d})^{1}\vert\vert{L^-}\rangle=-\frac{16n^2+(L-2)(L+3)}
{5(2n+1)}\sqrt{\frac{2L+1}{28n}}.}
\end{eqnarray}

\noindent (h)(2n+1,1)K=1 $L^- \rightarrow$ (2n+2,0)K=0 $(L+1)^+$
 transitions, e.g. $1^-\rightarrow2^+$,

\begin{eqnarray}
\lefteqn{\langle(L+1)^{+}\vert\vert(s^{\dagger}\tilde{p}+
p^{\dagger}\tilde{s})^{1}
\vert\vert{L^-}\rangle={-\frac{(2n-L+1)(2n+3L+6)}{(2n+1)}}\times}&&\nonumber \\
&&\sqrt{\frac{(2n+L+4)L}{60n(2n+3)}},
\end{eqnarray}
\begin{eqnarray}
\lefteqn{\langle(L+1)^{+}\vert\vert(d^{\dagger}\tilde{p}+
p^{\dagger}\tilde{d})^{1}
\vert\vert{L^-}\rangle={-\frac{4n^2-2(4L-1)n-3(L-1)(L+2)}{5(2n+1)}}
\times}&&\nonumber \\
&&\sqrt{\frac{(2n+L+4)L}{12n(2n+3)}},
\end{eqnarray}
\begin{eqnarray}
\lefteqn{\langle(L+1)^{+}\vert\vert(d^{\dagger}\tilde{f}+
f^{\dagger}\tilde{d})^{1}
\vert\vert{L^-}\rangle={-\frac{16n^2-12(L+1)n+3(L-1)(L+2)}{5(2n+1)}}
\times}&&\nonumber \\
&&\sqrt{\frac{(2n+L+4)L}{28n(2n+3)}}.
\end{eqnarray}

\noindent (i)(2n+1,1)K=1 $L^- \rightarrow$ (2n-2,2)K=0 $(L-1)^+$
 transitions, e.g. $1^-\rightarrow0^+$,

\begin{eqnarray}
\lefteqn{\langle(L-1)^{+}\vert\vert(s^{\dagger}\tilde{p}+
p^{\dagger}\tilde{s})^{1}
\vert\vert{L^-}\rangle={-\frac{2n-3L+3}{2n(2n+1)}}\times}&&\nonumber \\
&&\sqrt{\frac{(n+1)(2n+L+2)(L+1)\;\varphi(2n-1,L-1)}{15(2n+3)}},
\end{eqnarray}
\begin{eqnarray}
\lefteqn{\langle(L-1)^{+}\vert\vert(d^{\dagger}\tilde{p}+
p^{\dagger}\tilde{d})^{1}
\vert\vert{L^-}\rangle}&&\nonumber \\
&&={\frac{8n^3+12(2L+1)n^2-2(L-1)(2L-3)n-3(L-1)^2(L+3)}
{10n(2n+1)}\times}\nonumber \\
&&\sqrt{\frac{(n+1)(2n+L+2)(L+1)}{3(2n+3)\;\varphi(2n-1,L-1)}},
\end{eqnarray}
\begin{eqnarray}
\lefteqn{\langle(L-1)^{+}\vert\vert(d^{\dagger}\tilde{f}+
f^{\dagger}\tilde{d})^{1}
\vert\vert{L^-}\rangle}&&\nonumber \\
&&={\frac{32n^3-24(L-2)n^2-2(L-1)(3L-2)n+3(L-2)(L-1)^2}
{10n(2n+1)}\times}\nonumber \\
&&\sqrt{\frac{(n+1)(2n+L+2)(L+1)}{7(2n+3)\;\varphi(2n-1,L-1)}}.
\end{eqnarray}

\noindent (j)(2n+1,1)K=1 $L^- \rightarrow$ (2n-2,2)K=0 $L^+$
 transitions, e.g. $2^-\rightarrow2^+$,

\begin{eqnarray}
\lefteqn{\langle L^{+}\vert\vert(s^{\dagger}\tilde{p}+
p^{\dagger}\tilde{s})^{1}\vert\vert{L^-}\rangle={-\frac{1}{2n(2n+1)}}\times}
&&\nonumber \\&&\sqrt{\frac{(2n-L+2)(n+1)(2n+L+3)(2L+1)\;\varphi(2n-1,L)}{15}},
\end{eqnarray}
\begin{eqnarray}
\lefteqn{\langle L^{+}\vert\vert(d^{\dagger}\tilde{p}+
p^{\dagger}\tilde{d})^{1}\vert\vert{L^-}\rangle={\frac{4n^2+L^2+L-3}
{10n(2n+1)}}\times}&&\nonumber \\
&&\sqrt{\frac{(2n-L+2)(n+1)(2n+L+3)(2L+1)}{3\;\varphi(2n-1,L)}},
\end{eqnarray}
\begin{eqnarray}
\lefteqn{\langle L^{+}\vert\vert(d^{\dagger}\tilde{f}+
f^{\dagger}\tilde{d})^{1}\vert\vert{L^-}\rangle={\frac{16n^2-L^2-L-2}
{10n(2n+1)}}\times}&&\nonumber \\
&&\sqrt{\frac{(2n-L+2)(n+1)(2n+L+3)(2L+1)}{7\;\varphi(2n-1,L)}}.
\end{eqnarray}

\noindent (k)(2n+1,1)K=1 $L^- \rightarrow$ (2n-2,2)K=0 $(L+1)^+$
 transitions, e.g. $1^-\rightarrow2^+$,

\begin{eqnarray}
\lefteqn{\langle(L+1)^{+}\vert\vert(s^{\dagger}\tilde{p}+
p^{\dagger}\tilde{s})^{1}
\vert\vert{L^-}\rangle={-\frac{2n+3L+6}{2n(2n+1)}}\times}&&\nonumber \\
&&\sqrt{\frac{(n+1)(2n-L+1)L\;\varphi(2n-1,L+1)}{15(2n+3)}},
\end{eqnarray}
\begin{eqnarray}
\lefteqn{\langle(L+1)^{+}\vert\vert(d^{\dagger}\tilde{p}+
p^{\dagger}\tilde{d})^{1}
\vert\vert{L^-}\rangle}&&\nonumber \\
&&={\frac{8n^3-12(2L+1)n^2-2(L+2)(2L+5)n+3(L-2)(L+2)^2}
{10n(2n+1)}\times}\nonumber \\
&&\sqrt{\frac{(n+1)(2n-L+1)L}{3(2n+3)\;\varphi(2n-1,L+1)}},
\end{eqnarray}
\begin{eqnarray}
\lefteqn{\langle(L+1)^{+}\vert\vert(d^{\dagger}\tilde{f}+
f^{\dagger}\tilde{d})^{1}
\vert\vert{L^-}\rangle}&&\nonumber \\
&&={\frac{32n^3+24(L+3)n^2-2(L+2)(3L+5)n-3(L+2)^2(L+3)}
{10n(2n+1)}\times}\nonumber \\
&&\sqrt{\frac{(n+1)(2n-L+1)L}{7(2n+3)\;\varphi(2n-1,L+1)}}.
\end{eqnarray}

\noindent (l)(2n+1,1)K=1 $L^- \rightarrow$ (2n-2,2)K=2 $(L-1)^+$
 transitions, L is odd, \\ e.g. $3^-\rightarrow2^+$,

\begin{eqnarray}
\lefteqn{\langle(L-1)^{+}\vert\vert(s^{\dagger}\tilde{p}+
p^{\dagger}\tilde{s})^{1}
\vert\vert{L^-}\rangle=0,}
\end{eqnarray}
\begin{eqnarray}
\lefteqn{\langle(L-1)^{+}\vert\vert(d^{\dagger}\tilde{p}+
p^{\dagger}\tilde{d})^{1}
\vert\vert{L^-}\rangle={-\frac{1}{5}\sqrt{\frac{3(n+1)(2n+3)(L-2)(L-1)}
{2n(2n+1)L}}}\times}&&\nonumber \\
&&\sqrt{\frac{(2n-L+1)(2n+L)(2n+L+2)}
{\;\varphi(2n-1,L-1)}},
\end{eqnarray}
\begin{eqnarray}
\lefteqn{\langle(L-1)^{+}\vert\vert(d^{\dagger}\tilde{f}+
f^{\dagger}\tilde{d})^{1}
\vert\vert{L^-}\rangle={-\frac{2}{5}\sqrt{\frac{(n+1)(2n+3)(L-2)(L-1)}
{14n(2n+1)L}}}\times}&&\nonumber \\
&&\sqrt{\frac{(2n-L+1)(2n+L)(2n+L+2)}
{\;\varphi(2n-1,L-1)}}.
\end{eqnarray}

\noindent (m)(2n+1,1)K=1 $L^- \rightarrow$ (2n-2,2)K=2 $L^+$
 transitions, L is odd,\\  e.g. $3^-\rightarrow3^+$,

\begin{eqnarray}
\lefteqn{\langle L^{+}\vert\vert(s^{\dagger}\tilde{p}+
p^{\dagger}\tilde{s})^{1}
\vert\vert{L^-}\rangle=0,}
\end{eqnarray}
\begin{eqnarray}
\lefteqn{\langle L^{+}\vert\vert(d^{\dagger}\tilde{p}+
p^{\dagger}\tilde{d})^{1}
\vert\vert{L^-}\rangle={-\frac{1}{20n}\sqrt{\frac{6(n+1)(2n+3)(L-1)
(L+2)(2L+1)}
{(2n+1)L(L+1)}}}\times}&&\nonumber \\
&&\sqrt{(2n-L+1)(2n+L+2)},
\end{eqnarray}
\begin{eqnarray}
\lefteqn{\langle L^{+}\vert\vert(d^{\dagger}\tilde{f}+
f^{\dagger}\tilde{d})^{1}
\vert\vert{L^-}\rangle={-\frac{1}{10n}\sqrt{\frac{2(n+1)(2n+3)
(L-1)(L+2)(2L+1)}
{7(2n+1)L(L+1)}}}\times}&&\nonumber \\
&&\sqrt{(2n-L+1)(2n+L+2)}.
\end{eqnarray}

\noindent (n)(2n+1,1)K=1 $L^- \rightarrow$ (2n-2,2)K=2 $(L+1)^+$
 transitions, L is odd,\\  e.g. $1^-\rightarrow2^+$,

\begin{eqnarray}
\lefteqn{\langle(L+1)^{+}\vert\vert(s^{\dagger}\tilde{p}+
p^{\dagger}\tilde{s})^{1}
\vert\vert{L^-}\rangle=0,}
\end{eqnarray}
\begin{eqnarray}
\lefteqn{\langle(L+1)^{+}\vert\vert(d^{\dagger}\tilde{p}+
p^{\dagger}\tilde{d})^{1}
\vert\vert{L^-}\rangle={-\frac{1}{5}\sqrt{\frac{3(n+1)(2n+3)(L+2)(L+3)}
{2n(2n+1)(L+1)}}}\times}&&\nonumber \\
&&\sqrt{\frac{(2n-L-1)(2n-L+1)(2n+L+2)}
{\;\varphi(2n-1,L+1)}},
\end{eqnarray}
\begin{eqnarray}
\lefteqn{\langle(L+1)^{+}\vert\vert(d^{\dagger}\tilde{f}+
f^{\dagger}\tilde{d})^{1}
\vert\vert{L^-}\rangle={-\frac{2}{5}\sqrt{\frac{(n+1)(2n+3)(L+2)(L+3)}
{14n(2n+1)(L+1)}}}\times}&&\nonumber \\
&&\sqrt{\frac{(2n-L-1)(2n-L+1)(2n+L+2)}
{\;\varphi(2n-1,L+1)}}.
\end{eqnarray}

\noindent (o)(2n+1,1)K=1 $L^- \rightarrow$ (2n-2,2)K=2 $(L-1)^+$
 transitions, L is even,\\  e.g. $4^-\rightarrow3^+$,

\begin{eqnarray}
\lefteqn{\langle(L-1)^{+}\vert\vert(s^{\dagger}\tilde{p}+
p^{\dagger}\tilde{s})^{1}
\vert\vert{L^-}\rangle=0,}
\end{eqnarray}
\begin{eqnarray}
\lefteqn{\langle(L-1)^{+}\vert\vert(d^{\dagger}\tilde{p}+
p^{\dagger}\tilde{d})^{1}
\vert\vert{L^-}\rangle={\frac{1}{20n}\sqrt{\frac{6(n+1)(L-2)(L-1)}
{(2n+1)L}}}\times}&&\nonumber \\
&&\sqrt{(2n-L+2)(2n+L+1)(2n+L+3)},
\end{eqnarray}
\begin{eqnarray}
\lefteqn{\langle(L-1)^{+}\vert\vert(d^{\dagger}\tilde{f}+
f^{\dagger}\tilde{d})^{1}
\vert\vert{L^-}\rangle={\frac{1}{10n}\sqrt{\frac{2(n+1)(L-2)(L-1)}
{7(2n+1)L}}}\times}&&\nonumber \\
&&\sqrt{(2n-L+2)(2n+L+1)(2n+L+3)}.
\end{eqnarray}

\noindent (p)(2n+1,1)K=1 $L^- \rightarrow$ (2n-2,2)K=2 $L^+$
 transitions, L is even, \\ e.g. $2^-\rightarrow2^+$,

\begin{eqnarray}
\lefteqn{\langle L^{+}\vert\vert(s^{\dagger}\tilde{p}+
p^{\dagger}\tilde{s})^{1}
\vert\vert{L^-}\rangle=0,}
\end{eqnarray}
\begin{eqnarray}
\lefteqn{\langle L^{+}\vert\vert(d^{\dagger}\tilde{p}+
p^{\dagger}\tilde{d})^{1}
\vert\vert{L^-}\rangle={\frac{1}{5}\sqrt{\frac{3(n+1)(L-1)(L+2)(2L+1)}
{2n(2n+1)L(L+1)}}}\times}&&\nonumber \\
&&\sqrt{\frac{(2n-L)(2n-L+2)(2n+L+1)(2n+L+3)}
{\;\varphi(2n-1,L)}},
\end{eqnarray}
\begin{eqnarray}
\lefteqn{\langle L^{+}\vert\vert(d^{\dagger}\tilde{f}+
f^{\dagger}\tilde{d})^{1}
\vert\vert{L^-}\rangle={\frac{2}{5}\sqrt{\frac{(n+1)(L-1)(L+2)(2L+1)}
{14n(2n+1)L(L+1)}}}\times}&&\nonumber \\
&&\sqrt{\frac{(2n-L)(2n-L+2)(2n+L+1)(2n+L+3)}
{\;\varphi(2n-1,L)}}.
\end{eqnarray}

\noindent (q)(2n+1,1)K=1 $L^- \rightarrow$ (2n-2,2)K=2 $(L+1)^+$
 transitions, L is even, \\ e.g. $2^-\rightarrow3^+$,

\begin{eqnarray}
\lefteqn{\langle(L+1)^{+}\vert\vert(s^{\dagger}\tilde{p}+
p^{\dagger}\tilde{s})^{1}
\vert\vert{L^-}\rangle=0,}
\end{eqnarray}
\begin{eqnarray}
\lefteqn{\langle(L+1)^{+}\vert\vert(d^{\dagger}\tilde{p}+
p^{\dagger}\tilde{d})^{1}
\vert\vert{L^-}\rangle={\frac{1}{20n}\sqrt{\frac{6(n+1)(L+2)(L+3)}
{(2n+1)(L+1)}}}\times}&&\nonumber \\
&&\sqrt{(2n-L)(2n-L+2)(2n+L+3)},
\end{eqnarray}
\begin{eqnarray}
\lefteqn{\langle(L+1)^{+}\vert\vert(d^{\dagger}\tilde{f}+
f^{\dagger}\tilde{d})^{1}
\vert\vert{L^-}\rangle={\frac{1}{10n}\sqrt{\frac{2(n+1)(L+2)(L+3)}
{7(2n+1)(L+1)}}}\times}&&\nonumber \\
&&\sqrt{(2n-L)(2n-L+2)(2n+L+3)}.
\end{eqnarray}

\subsection{The M1 transition matrix elements}
\label{s2s3}

\noindent (a)(2n+1,1)K=1 $L^- \rightarrow$ (2n+3,0)K=0 $(L-1)^-$
 transitions, e.g. $2^-\rightarrow1^-$,

\begin{eqnarray}
\lefteqn{\langle(L-1)^{-}\vert\vert(p^{\dagger}\tilde{p})^{1}
\vert\vert{L^-}\rangle={-\frac{(2n-L+2)(2n+L+3)}{5(2n+1)}}\times}&&\nonumber \\
&&\sqrt{\frac{3(2n-L+4)(L+1)}{8n(n+1)(2n+3)}},
\end{eqnarray}
\begin{eqnarray}
\lefteqn{\langle(L-1)^{-}\vert\vert(f^{\dagger}\tilde{f})^{1}
\vert\vert{L^-}\rangle={-\frac{16n^2+(L-2)(L+3)}{10(2n+1)}}\times}
&&\nonumber \\
&&\sqrt{\frac{3(2n-L+4)(L+1)}{28n(n+1)(2n+3)}},
\end{eqnarray}
\begin{eqnarray}
\lefteqn{\langle(L-1)^{-}\vert\vert(d^{\dagger}\tilde{d})^{1}
\vert\vert{L^-}\rangle=\sqrt{\frac{3n(2n-L+4)(L+1)}{10(n+1)(2n+3)}}.}
\end{eqnarray}

\noindent (b)(2n+1,1)K=1 $L^- \rightarrow$ (2n+3,0)K=0 $(L+1)^-$
 transitions, e.g. $2^-\rightarrow3^-$,

\begin{eqnarray}
\lefteqn{\langle(L+1)^{-}\vert\vert(p^{\dagger}\tilde{p})^{1}
\vert\vert{L^-}\rangle={-\frac{(2n-L+2)(2n+L+3)}{5(2n+1)}}\times}&&\nonumber \\
&&\sqrt{\frac{3(2n+L+5)L}{8n(n+1)(2n+3)}},
\end{eqnarray}
\begin{eqnarray}
\lefteqn{\langle(L+1)^{-}\vert\vert(f^{\dagger}\tilde{f})^{1}
\vert\vert{L^-}\rangle={-\frac{16n^2+(L-2)(L+3)}{10(2n+1)}}\times}&&\nonumber \\
&&\sqrt{\frac{3(2n+L+5)L}{28n(n+1)(2n+3)}},
\end{eqnarray}
\begin{eqnarray}
\lefteqn{\langle(L+1)^{-}\vert\vert(d^{\dagger}\tilde{d})^{1}
\vert\vert{L^-}\rangle=\sqrt{\frac{3n(2n+L+5)L}{10(n+1)(2n+3)}}.}
\end{eqnarray}
\vspace{1.0cm}

\section{Studies of octupole vibration in $^{152}$Sm}
\label{s3}

\subsection{The spectrum}
\label{s3s1}

\begin{figure}
\begin{center}
\epsfig{figure=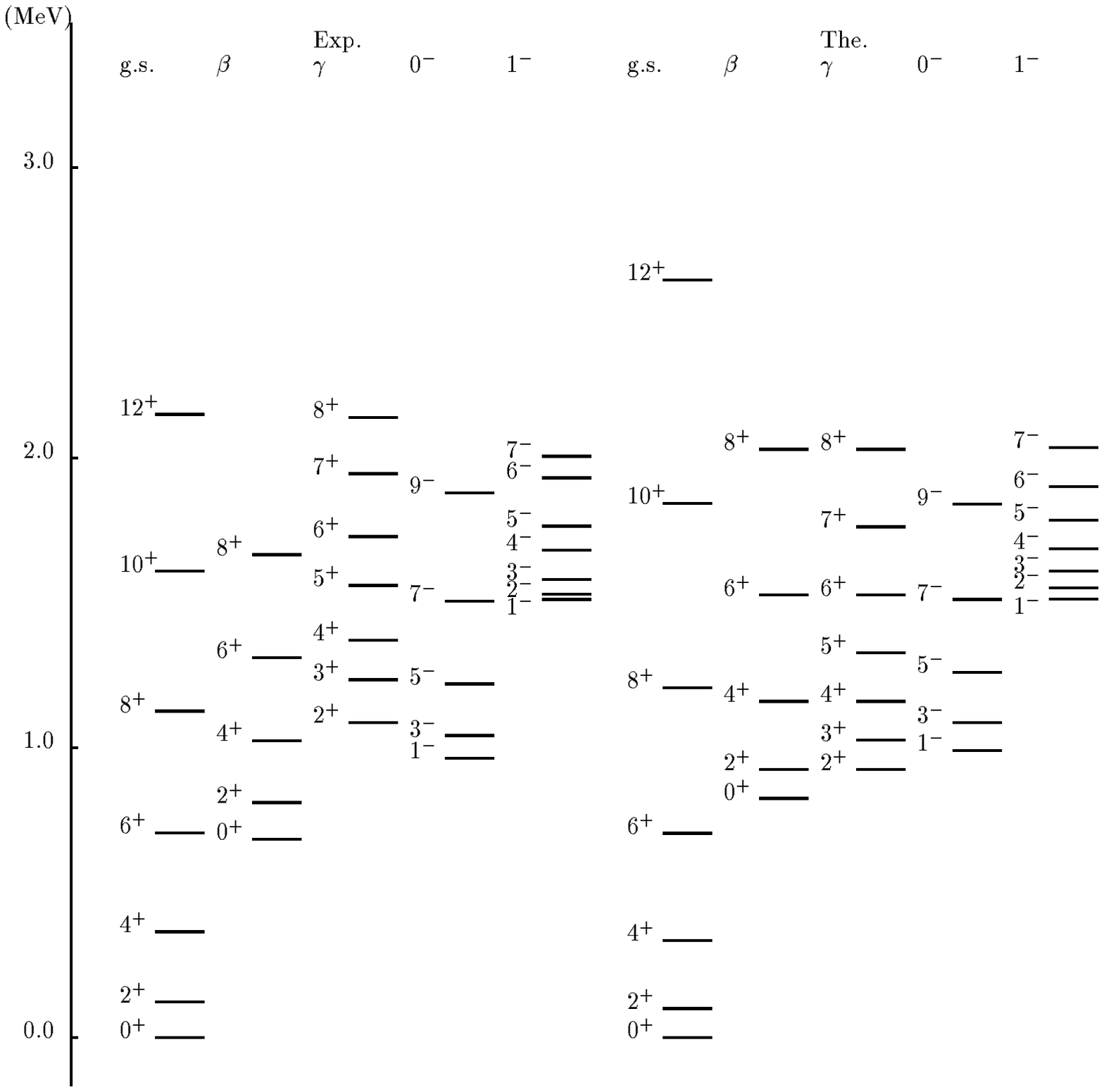,width=15cm,height=15cm}
\end{center}
\caption{Spectra of $^{152}$Sm.}
\label{f1}
\end{figure}

The spectrum of the spdf SU(3) is 
compared with data\cite{r29} in Fig.\ref{f1}. 
The parameters are set as: $a_1$=-7.49keV,$\epsilon_-$=4.65MeV,
$a_3$=-8.90keV. And $a_4$ is 16.75keV for the positive parity 
states and 9.67keV for the negative parity states, respectively. 
The smaller value of $a_4$ for the negative parity states reflects 
an increase of moment of inertia for the negative parity states. 
This phenomenon has also been found in uranium nuclei\cite{r30}.

The general agreement between experiment and calculation is quite good. 
The five bands(3 with positive parity and 2 with negative parity) are 
all well reproduced. However, as expected, due to the approximate 
nature of the SU(3) dynamic symmetry in this nucleus, there are 
discrepancies between them, for instance the degeneracy of $\beta$ 
and $\gamma$ bands is broken in experiment. 
These detailed structures should be 
the task of an elaborate numerical studies. Here we are content with 
the result that the gross structure of the nucleus can be accounted for
well by the SU(3) dynamic symmetry.

\subsection{The E1 transition rates}
\label{s3s2}

We have applied the results in Sect. 2 and those in Ref.\cite{r23} to 
$^{152}$Sm with negative parity states. As for the transition operator,
simplicity can be obtained if one choose the transition operator as some
generator of the dynamical group. In the sdpf-IBM, there is no SU(3)
generator that can be used as E1 transition operator. However there is
an O(10) group generator\cite{r13} that can change the parity and can be used as
E1 transition operator. However this should not be taken to seriously
because the nuclear hamiltonian is a ``residual" strong interaction, and
the E1 transition is induced by electromagnetic interaction. There is
no strong argument that the operator in the hamiltonian and the
operator in the transition should have the same form. With our present
knowledge, the transition operator should be determined from
experimental data.  Since a generator form transition operator can
sometimes bring simplicity, it is useful to explore if the O(10)
generator can achieve some simplification. 
At first we take the transition operator as the O(10) generator:
\begin{eqnarray}
T(E1)^1_\mu=e_1D^1_\mu,
\end{eqnarray}
where
\begin{eqnarray}
D^1_\mu=\sqrt{\frac{1}{2}}(s^{\dagger}\tilde{p}+
p^{\dagger}\tilde{s})^{1}_\mu-\sqrt{\frac{4}{5}}(p^{\dagger}\tilde{d}+
d^{\dagger}\tilde{p})^{1}_\mu+\sqrt{\frac{7}{10}}(d^{\dagger}\tilde{f}+
f^{\dagger}\tilde{d})^{1}_\mu.
\end{eqnarray}
The calculation values are listed in the column 
labeled Cal.1 in Table \ref{t1}.
The agreement between experiment and calculation is very good, 
in particular for those 
transition from $0^-$-band to the positive parity states.
The effect of the different terms in the transition operator may play 
different roles, a combination of them may conceal some of the properties
of the individual term. For instance, in the SU(3) limit of the sdg-IBM,
each term in the E2 transition operator has an L(L+3) dependence, which 
plays an important role at large L. But this dependence disappears for 
the SU(3) generator form of the E2 transition operator; this L(L+3) 
dependence term is concealed, and leads to the reduction of collectivity
problem\cite{r30a,r31}. To see the effect of each individual term,
we have calculated the reduced matrix elements of 
sp, dp and df terms respectively. The calculation results of dp term 
alone are listed in the column labeled Cal.2 in Table \ref{t1}. 
We found that 
the experiment could be reproduced well by using solely 
the dp term in $^{152}$Sm. 
This indicates that the dp term plays a leading role in the E1 
transition in the low-lying states of $^{152}$Sm. The inclusion 
of p-boson for the low-lying negative parity states is very important,
 as has been pointed in Ref.\cite{r9,r10,r11,r11a}.

Finally, we used numerical fitting to improve the agreement. 
The general E1 transition operator is:
\begin{eqnarray}
T(E1)^1_\mu=e_1[(s^{\dagger}\tilde{p}+
p^{\dagger}\tilde{s})^{1}_\mu+\chi_{dp}(p^{\dagger}\tilde{d}+
d^{\dagger}\tilde{p})^{1}_\mu+\chi_{df}(d^{\dagger}\tilde{f}+
f^{\dagger}\tilde{d})^{1}_\mu].
\end{eqnarray}
Since we have fixed the coefficient of sp term to 1, we can 
evaluate the relative weights of these three terms through 
the values of $\chi_{dp}$ and $\chi_{df}$. The results have 
been listed in the column labeled Cal.3 in able 1, and the 
parameters are $\chi_{dp}=81.669$, $\chi_{df}=-4.975$. As 
expected, the value of $\chi_{dp}$ evinces that dp term is 
far more important than sp and dp terms.

\section{Applications to other deformed rare-earth nuclei}
\label{s4}

 We have expanded our work to other deformed nuclei,
namely $^{154}$Sm, $^{156-160}$Gd, $^{160-162}$Dy, where E1 experimental
data are available.
 The results are shown in Table\ref{t3} together with the experimental
 data\cite{r32,r33,r34,r35,r36,r37}.
The arrangement of Table \ref{t3} is the same as that in Table \ref{t1}.
The parameters $e_1$,
$\chi_{dp}$ and $\chi_{df}$ are listed in Table \ref{t4}. 
The spectra of 
the nuclei considered here can be well described by the SU(3) dynamic 
symmetry. Moreover, the E2 transition rates of the positive parity
states can also be reasonably well 
described by the SU(3) dynamic symmetry\cite{r38}.

From Table \ref{t3}, 
we can find that the B(E1) values from $0^-$ band to ground state
band in the nuclei studied are all well reproduced.
In particular, it is noted that the ratio $B(E1,1^-\rightarrow 2^+_{gs})/
B(E1,1^-\rightarrow 0^+_{gs})$ is approximately 2 in the spdf IBM SU(3)
octupole vibration limit. There is because there is a factor $\sqrt{L}$
in the E1 transition matrix element for $L^-\rightarrow (L-1)^+$, and a
factor of $\sqrt{L+1}$ in the E1 transition matrix element
$L^-\rightarrow (L+1)^+$. The other factors in the two matrix elements are
almost the same for $N\approx 10$. When $L=1$, it gives a ratio of 2 for
the two B(E1) values. This SU(3) octupole vibration property is all
satisfied by the 7 nuclei studied in this work. This property can be
used to extract the $K$ values for the low-lying $1^-$ state  of
deformed nuclei. For the $K^{\pi}=1^-$ bandhead, the transition B(E1) to
ground state band states is just the opposite, where the transition to
$0^+_{gs}$ state is stronger than to $2^+_{gs}$ state.

In $^{160}$Dy, it is noted that the experimental B(E1) values vary over
a wide range from 6.8$\times 10^{-3}$ W.u. to 3.10$\times 10^{-8}$ W.u.,
an order of 5 of change! If one fits the transition from $K^{\pi}=0^-$ band,
then the transition from $K^{\pi}=1^-$ 
band will be too large compared with
experimental data. Conversely if one tries to reproduce the transitions
from $K^{\pi}=1^-$ band, then the transitions from 
$K^{\pi}=0^-$ band will be too small.
The same situation occurred also in Ref.\cite{r19}. We therefore suggest
that the origins of the $1^-_2$ state which is the bandhead of the 
$K^{\pi}=0^-$-band, 
and $2^-_1$ state which is a member of the $K^{\pi}=1^-$-band are
different. They can not be described simultaneously in the spdf IBM. We
also noticed that the assignment of the $1^-_2$ state as a member of
$K^{\pi}=0^-$-band is only temporary in experiment\cite{r19,r37}.
If we consider only the transitions
from $2^-_1$ to positive parity states,  agreement is obtained. 

In $^{158}$Gd, there are 7 B(E1) experimental data. The B(E1) values
vary over also a wide range from 0.00633 W.u. to 1.21$\times 10^{-5}$ W.u..
They are all well reproduced by the spdf IBM SU(3) octupole vibration
limit. To our present experimental knowledge,
we can say that $^{158}$Gd is the best nucleus showing
the SU(3) octupole vibration dynamical symmetry. We expect that with the
development in experiment, other nuclei with this dynamical symmetry 
will be
discovered in other mass regions, for instance in the actinide region.

From Table \ref{t4}, 
we can find that in other nuclei:(a). both dp and df terms play 
important roles in these transitions, so we can not acquire the 
good agreement only by dp term alone. (b). the values 
of $\chi_{dp}$ and $\chi_{df}$ change in a relative narrow range,
 and dp and df terms have almost the same importance in these nuclei, 
which is different from $^{152}$Sm. And the signs of $\chi_{dp}$ and 
$\chi_{df}$ remain invariant in a given isotope chain. 
(c). in the isotopes of Gd, the 
changes of $\chi_{dp}$ and $\chi_{df}$ obey the following regularity: 
with the number of neutron increasing, $\chi_{df}$ decreases, however, 
the absolute value of $\chi_{dp}$ increases first, then decreases. 
 In $^{156}$Gd, the absolute value of $\chi_{df}$
(4.37) is greater than that of $\chi_{dp}$(-1.62), and in $^{160}$Gd, 
the absolute value of $\chi_{df}$(1.61) is less than that of 
$\chi_{dp}$(-2.08), which implies that dp and df terms may play 
different roles in different nuclei. In $^{158}$Gd, the absolute 
value of $\chi_{dp}$ reaches its maximum.It requires more experimental data in  
relevant nuclei to check this relationship between the value of 
$\chi_{dp}$ and the number of neutron in a given isotope chain.

From the E1 transition formulae, we found that 
the E1 transition between $0^-$-band and $\gamma$-band is zero in 
the SU(3) limit. However, such transitions occurred in $^{152}$Sm, 
$^{160}$Dy. In $^{152}$Sm, $B(E1;4^+_{\gamma}\rightarrow3^-_1)=3.2
\times10^{-5}(13)$, this value is very small. A small symmetry-breaking
will give a nonzero value for this transition. 
For instance if we introduce some mixing between the $\beta$ and
$\gamma$ band, that is,
\begin{eqnarray}
\vert\beta\rangle^{'}=a_1\vert\beta\rangle+a_2\vert\gamma\rangle,\\
\vert\gamma\rangle^{'}=-a_2\vert\beta\rangle+a_1\vert\gamma\rangle,
\end{eqnarray}
where the apostrophe labeled the new $\beta$-band and $\gamma$-band. 
When $\vert{a_2}\vert=0.0118$, $B(E1;4^+_{\gamma}\rightarrow3^-_1)=3.2
\times10^{-5}$, and because $\vert{a_2}\vert$ is so small that other 
properties of $\gamma$-band could not be disturbed much. 
However, in 
$^{160}$Dy, $B(E1;1^-_1\rightarrow2^+_{\gamma})=0.018(18)$, a very large
value. A simple symmetry breaking can not solve this problem.
This is another reason that the $1^-_2$ state in $^{160}$Dy may not
belong to the $K^{\pi}=0^-$ band.
In $^{160}$Dy, the E1 transition between $3^-_3$ of possible 
$K^{\pi}=0^-$-band to $\gamma$ band are also observed, which confirmed us that 
this band is not likely the $0^-$-band.

Besides absolute B(E1) data, we also compare relative intensities.
If one assumes that E1 is dominant in parity changing transitions
and ignore higher order transitions, then the transition probabilities
and
 hence the intensities are
\begin{eqnarray}
int\propto E^3_{\gamma}B(E1).
\end{eqnarray}
Using the values of $\chi_{dp}$ and $\chi_{df}$ obtained in the 
previous work for the B(E1) value, we have calculated the relative 
intensities in $^{152,154}$Sm\cite{r29,r32}, 
$^{156,158,160}$Gd\cite{r33,r34,r35}, 
$^{160,162}$Dy\cite{r35,r36}. 
The experimental data are taken from the 
references after each nucleus and from Ref.\cite{r19,r37}. The calculated 
results are shown in Table \ref{t5} with experimental data.

From Table \ref{t5}, we see
besides general agreement,
 there exist serious deviations in $^{152,154}$Sm: in 
experiment, the relative intensity from $1^-$ of $K^{\pi}=1^-$-band to 
$2^+_{gs}$ is much greater than that to $0^+_{gs}$, 
but in calculation, the situation is reversed.
In general, the square of reduced matrix element for
transition from $1^-$ state of the $K^{\pi}=1^-$-band to $0^+_{gs}$ is almost 
as twice as that to $2^+_{gs}$. 
The calculated ratio of intensities should be of the order of 1.
However we noticed that in $^{152}$Sm, the transition to $2_{gs}^+$
also involves M2 transitions. In the calculation we have just included
the E1 contribution. In $^{154}$Sm, the identification of $1^-$ state is
only tentative. Experimental measurement of the E1 transition rate from
$1^-$ to $2^+_{gs}$ will be very useful for checking this calculation,
since in our model there is little freedom to adjust this transition
rate. 
 
\section{Summary}
\label{s5}

We have given analytical expressions for E1 and M1 transitions involving
 low-lying negative parity states in the SU(3) limit. These E1
transition
 formulae plus those given in Ref.\cite{r23} 
exhaust the whole possibilities of
E1 transitions from $0^-$ and $1^-$ band to the ground state band, the
beta and gamma band.
We also applied these analytical results to  to some
deformed rare-earth nuclei to check the SU(3) octupole vibration 
prediction. Though there have been some deviation, the main features
are checked by the experimental data. Some of the discrepancies are
expected because of the approximate nature of the dynamical symmetry in
real nucleus, for instance the breaking of the beta and gamma band
states degeneracies, some small transitions between gamma band and 
$K^{\pi}=0^-$
band. They can be resolved by symmetry breaking. 
Some of the discrepancies can not be nailed down at the moment due to
lack of experimental data, for instance, the 
$K{\pi}=0^-$ band identification
in $^{160}$Dy. Another important issue which remain to be checked by
experiment is the
anomalous intensity ratio of $K^{\pi}=1^-$ bandhead in $^{152,154}$Sm.
Comparing with the sd-f IBM studies of Cottle and Zamfir\cite{r19}, 
we see that the
agreements of the calculations of the two models 
with experimental data are of the
same quality. 
In this spdf IBM study, the p boson is very important 
in these deformed nuclei, 
even critical in some nucleus, i.e. $^{152}$Sm.
In the sd-f IBM study, the f boson is most important.
 From these empirical studies alone, it is not conclusive
which model is more appropriate. However, the spdf IBM is more general,
and sd-f IBM is only a special case of the spdf IBM. The two
calculations can be thought as being two different parameterizations of
the spdf IBM. In the sd-f IBM, it can be considered as a special case in
which the energy of the p boson is put infinitely high. In the SU(3)
octupole vibration limit,  the energy of the p boson is put equal to the
energy of the f boson. 
However in view of the simple mathematical structure of spdf IBM, 
the microscopic studies\cite{r11a} and the findings by
Kusnezov\cite{kus}, the spdf interacting boson model is one promising
model for describing negative parity collective states in even-even nuclei.

\section{Acknowledgments}
The authors acknowledge the support of China National Natural Science
Foundation, China State Education Ministry, Fok Ying Tung Education
foundation. EG Zhao also acknowledges the support of Advanced Visitors
Scheme of Tsinghua University. The authors also thank the referee for
expert reading and helpful criticism of the manuscript.

\vskip 2cm
\noindent Corrigenda

In table 1 of reference \cite{r23}, the lines concerning $^{156}$Gd
should be replaced by

\begin{center}
\begin{tabular}{lllcc}\hline
Nucleus    & $I_i$   & $I_f$   & $B(E1,I_i\rightarrow I_f)_{cal}(W.u.)$ 
& $B(E1,I_i\rightarrow I_f)_{exp}(W.u.)$ \\ \hline
$^{156}$Gd & $1^+_2$ & $0^+_1$ & 0.0025 & 0.0025(14) \\
           &         & $2^+_1$ & 0.005  & 0.006(3) \\ \hline
\end{tabular}
\end{center}

\begin{table}
\begin{center}
\begin{tabular}{c l l l l l l }\hline
 $K^{\pi}$  &  $I_i$ &   $ I_f$         & Cal.1(W.u.) & Cal.2(W.u.)&
 Cal.3(W.u.)&  Exp.(W.u.)  \\ \hline
  $0^-$     &$1^-_1$ &   $0^+_{gs}$     & 0.0042      &  0.0042    &     0.0037    &  0.0042(4)   \\ 
  $0^-$     &$1^-_1$ &   $2^+_{gs}$     & 0.0081      &  0.00118   &       0.0099    &  0.0077(7)   \\
  $0^-$     &$1^-_1$ &   $2^+_{\beta}$  & 1.4$\times10^{-5}$       &
  4.3$\times10^{-3}$ &4.0$\times10^{-3}$& 1.3(4)$\times10^{-4}$  \\
  $0^-$     &$3^-_1$ &   $2^+_{gs}$     & 0.0055      &  0.0044    &     0.0039    &  0.0081(16)  \\
  $0^-$     &$3^-_1$ &   $4^+_{gs}$     & 0.0067      &  0.00119    &     0.0099    &  0.0082(16)  \\
  $1^-$     &$2^-_1$ &   $2^+_{gs}$     & 0.0015      &  0.0015    &     0.0010    &  0.0027     \\
  $1^-$     &$2^-_1$ &   $2^+_{\beta}$  & 3.8$\times10^{-4}$       &
6.6$\times10^{-4}$   &3.3$\times10^{-4}$&5.6$\times10^{-5}$      \\
  $1^-$     &$2^-_1$ &   $2^+_{\gamma}$ & 0.0002      &  0.0042    & 
  0.0033    &  0.011      \\
  $1^-$     &$2^-_1$ &   $3^+_{\gamma}$ & 0.00035     &  0.0072    &     0.0056    &  0.0061  \\   \hline
\end{tabular}
\end{center}
\caption{B(E1) values in $^{152}$Sm}
\label{t1}
\end{table}

\begin{table}
\begin{center}
\begin{tabular}{c l l l l l l l}\hline
  Nucleus    &  $K^{\pi}$  & $I_i$  &   $ I_f$     &Cal.1(W.u.) &Cal.2(W.u.) &  Cal.3(W.u.) & 
Exp.(W.u.)\\ \hline
  $^{154}$Sm & $0^-$       &$1^-_1$ &   $0^+_{gs}$ & 0.0062     &  0.0062   &  0.0041      &0.0062(4)   \\
             & $0^-$       &$1^-_1$ &   $2^+_{gs}$ & 0.0120     &  0.0163   &  0.0106      & 0.0117(26) \\
             & $0^-$       &$3^-_1$ &   $2^+_{gs}$ & 0.0081     &  0.0064   &  0.0045      &0.0078(10)  \\
             & $0^-$       &$3^-_1$ &   $4^+_{gs}$ & 0.0099     &  0.0162   &  0.0104      &0.0093(14)  \\
  $^{156}$Gd & $1^-$       &$1^-_1$ &   $0^+_{gs}$ & 0.0005     &  0.0007   &  0.0018      &0.0016(12)  \\
             & $1^-$       &$1^-_1$ &   $2^+_{gs}$ & 0.0004     &  0.0001   &  0.0009      &0.0020(15)  \\
             & $1^-$       &$1^-_1$ & $0^+_{\beta}$& 0.00015    &  0.00040  &  0.00020     &0.00030(24) \\
             & $0^-$       &$1^-_2$ &   $0^+_{gs}$ & 0.0025     &  0.0025   &  0.0029      &0.0025(14)  \\
             & $0^-$       &$1^-_2$ &   $2^+_{gs}$ & 0.005      &  0.006    &  0.0051      &0.006(3)    \\
             & $0^-$       &$1^-_2$ & $2^+_{\beta}$&9$\times10^{-6}$&0.0025 &  0.0004      &0.0007(5)   \\
  $^{158}$Gd & $0^-$       &$1^-_2$ &   $0^+_{gs}$ & 0.0035     & 0.0035    &  0.0028      &0.0035(8)   \\
             & $0^-$       &$1^-_2$ &   $2^+_{gs}$ & 0.0068     & 0.0088    &  0.0056      &0.0063(16)  \\
             & $1^-$       &$3^-_1$ &   $2^+_{gs}$ & 0.00047    & 0.00105   &  0.00029     &0.00033(10) \\
             & $1^-$       &$3^-_1$ &   $4^+_{gs}$ & 0.00088    & 0.00009   &  0.00041     &0.00029(8)  \\
             & $1^-$       &$2^+_{\beta}$& $1^-_1$ & 6.2$\times10^{-5}$
&  2.6$\times10^{-5}$      &1.3 $\times10^{-5}$    & 6.4(8)$\times10^{-5}$
\\
             & $1^-$       &$2^+_{\beta}$& $2^-_1$ & 3.16$\times10^{-5}$
&  5.00$\times10^{-4}$      &1.19$\times10^{-5}$    & 1.21(5)$\times10^{-5}$ \\
             & $1^-$       &$2^+_{\beta}$& $3^-_1$ & 2.54$\times10^{-4}$
&  1.22$\times10^{-3}$     &1.11$\times10^{-5}$    & 1.89$\times10^{-4}$(24)
\\
  $^{160}$Gd & $0^-$       &$1^-_1$ &   $0^+_{gs}$ & 0.0032     & 0.0032    &  0.0021      &0.0032(9)   \\
             & $0^-$       &$1^-_1$ &   $2^+_{gs}$ & 0.0062     & 0.0079  
&  0.0041      &0.0060(17)  \\
             & $0^-$       &$3^-_1$ &   $2^+_{gs}$ & 0.0042     & 0.0035    &  0.0026      &0.0016(5)   \\
             & $0^-$       &$3^-_1$ &   $4^+_{gs}$ & 0.0052     &  0.0077   &  0.0035      &0.0013(4)   \\
  $^{160}$Dy & $1^-$       &$2^-_1$ &   $2^+_{gs}$ & 1.4$\times10^{-6}$     &  1.2$\times10^{-7}$      &3.10$\times10^{-8}$    & 3.10(17)$\times10^{-8}$
 \\
             & $1^-$       &$2^-_1$&   $2^+_{\gamma}$&1.8$\times10^{-7}$    &  3.4$\times10^{-7}$      &1.89$\times10^{-7}$    & 1.89(10)$\times10^{-7}$ 
\\
             & $1^-$       &$2^-_1$&   $3^+_{\gamma}$&3.0$\times10^{-7}$    & 6.1$\times10^{-7}$       &2.49$\times10^{-7}$    & 2.49(13)$\times10^{-7}$
\\
             & $0^-$       &$1^-_2$&   $0^+_{gs}$  & 3.8$\times10^{-6}$     & 3.8$\times10^{-7}$       &7.3$\times10^{-7}$     & 3.8(4)$\times10^{-3}$
\\
             & $0^-$       &$1^-_2$&   $2^+_{gs}$  & 7.4$\times10^{-6}$     & 9.4$\times10^{-7}$       &1.5$\times10^{-6}$     & 6.8(5)$\times10^{-3}$ 
\\
  $^{162}$Dy & $0^-$       &$1^-_1$ &   $0^+_{gs}$ & 0.0026       & 0.0026  &  0.0026      &0.0026(4)   \\
             & $0^-$       &$1^-_1$ &   $2^+_{gs}$ & 0.0051       & 0.0064    
&  0.0060    &0.0060(19) \\ \hline
\end{tabular}
\end{center}
\caption{B(E1) values in deformed rare-earth nuclei}
\label{t3}
\end{table}

\begin{table}
\begin{center}
\begin{tabular}{c l l l  }\hline
 Nucleus    &   $e_1$  &  $\chi_{dp}$  & $\chi_{df}$ \\ \hline
 $^{152}$Sm &  0.00101 & 81.67       & $-4.98$      \\
 $^{154}$Sm &  0.00097 & 83.15       & $-8.59$       \\
 $^{156}$Gd &  0.0118  & $-1.62$     & 4.37       \\
 $^{158}$Gd &  0.0136  & $-3.83$     & 3.68       \\
 $^{160}$Gd &  0.0120  & $-2.08$     & 1.61       \\
 $^{160}$Dy &  0.0007  & $-0.49$     & $-0.56$      \\
 $^{162}$Dy &  0.0417  & $-0.75$    & $-0.35$   \\   \hline
\end{tabular}
\end{center}
\caption{Parameters in E1 transition operator. The unit of $e_1$ is in
0.28389 $A^{1/3}$ $e$ $fm$, which gives the B(E1) in Weissknopf unit.}
\label{t4}
\end{table}

\begin{table}
\begin{center}
\begin{tabular}{c l l l l l l }\hline
  Nucleus    &  $E_{level}(keV)$  & $K^{\pi}$  & $I_i$  &   $ I_f$ 
&   Cal.  &   Exp.   \\ \hline
  $^{152}$Sm &   963              &  $0^-$     &$1^-_1$ &   $0^+_{gs}$
&   55.6  &  82.3(7) \\
             &                    &            &        &   $2^+_{gs}$
&  100    &  100(2)  \\
             &                    &            &        &   $2^+_{\beta}$
& 0.24    &  0.010(3)\\
             &  1041              &            &$3^-_1$ &   $2^+_{gs}$
&  98.2   &  100(4)  \\
             &                    &            &        &   $4^+_{gs}$
&  100    &  40.4(19)\\
             &  1221              &            &$5^-_1$ &   $4^+_{gs}$
&  100    &  100(3)  \\
             &                    &            &        &   $6^+_{gs}$
&  72     &  24(3)   \\
             &  1505              &            &$7^-_1$ &   $6^+_{gs}$
&  100    &  100(3)  \\
             &                    &            &        &   $8^+_{gs}$
&  51     &  12      \\
             &  1511              &  $1^-$     &$1^-_2$ &   $0^+_{gs}$
&  100    &  0.85(6) \\
             &                    &            &        &   $2^+_{gs}$
&  13.3   &  100(3)  \\
             &                    &            &        &   $0^+_{\beta}$
& 8.38    &  0.09(5) \\
             &                    &            &        &   $2^+_{\beta}$
& 0.035   &  1.42(25)\\
             &  1530              &            &$2^-_1$ &   $2^+_{gs}$
&  100    &  100(11) \\
             &                    &            &        &   $2^+_{\beta}$
& 4.55    &  0.28(4) \\
             &                    &            &        &   $2^+_{\gamma}$
&  10.50  & 13.43(9) \\
             &                    &            &        &   $3^+_{\gamma}$
&  5.283  & 2.127(24)\\
             &  1579              &            &$3^-_2$ &   $2^+_{gs}$
& 100     & 35.3(3)  \\
             &                    &            &        &   $4^+_{gs}$
&  1.15   &  100(5)  \\
             &                    &            &        &   $2^+_{\beta}$
& 14.4    &  6.6(3)  \\
             &                    &            &        &   $4^+_{\beta}$
& 0.289   &  1.30(9) \\
             &                    &            &        &   $2^+_{\gamma}$
&  2.11   & 2.07(15) \\
             &                    &            &        &   $4^+_{\gamma}$
&  1.2    & 0.41(5)  \\
             &  1764              &            &$5^-_2$ &   $4^+_{gs}$
&  40     & 67(34)   \\
             &                    &            &        &   $6^+_{gs}$
&  100    &  100(33) \\
             &  1879              &  $0^-$     &$9^-_1$ &   $8^+_{gs}$
&  100    &  100(3)  \\
             &                    &            &        &   $10^+_{gs}$
&  31     &  15      \\
             &  1930              &   $1^-$    &$6^-_1$ &   $6^+_{gs}$
&  100    &  82(20)  \\
             &                    &            &        &   $5^+_{\gamma}$
& 5.4     &  100(4)  \\
             &  2004              &            &$7^-_2$ &   $6^+_{gs}$
&  100    &  100(7) \\
             &                    &            &        &   $8^+_{gs}$
&  14     &  71(24)  \\
             &  2291              &            &$9^-_2$ &   $8^+_{gs}$
&  100    &  100(5)   \\
             &                    &            &        &   $10^+_{gs}$
&  32     &  27(3)  \\
             &                    &            &        &   $8^+_{\beta}$
&  58     &  13(4) \\
             &  2641              &            &$11^-_1$&   $10^+_{gs}$
&  100    & 100(5)   \\
             &                    &            &        &   $12^+_{gs}$
&  62     & 59(24) \\ 
             &                    &            &        &   $10^+_{\beta}$
&  72     & 24 \\

  $^{154}$Sm &   921              &  $0^-$     &$1^-_1$ &   $0^+_{gs}$
&   52    &  65(2)   \\
             &                    &            &        &   $2^+_{gs}$
&  100    &  100(2)  \\
             &  1012              &            &$3^-_1$ &   $2^+_{gs}$
&  83    &  100(2)  \\
             &                    &            &        &   $4^+_{gs}$
&  100    &  60(1)   \\
             &  1181              &            &$5^-_1$ &   $4^+_{gs}$
&  100    &  100(3)  \\
             &                    &            &        &   $6^+_{gs}$
&   98  &  29.0(7) \\
             &  1476              &  $1^-$     &$1^-_2$ &   $0^+_{gs}$
&  100    &  0.5     \\
             &                    &            &        &   $2^+_{gs}$
&  15     &  100     \\
             &  1515              &            &$2^-_1$ &   $2^+_{gs}$
&  100    &  100(19) \\
             &                    &            &        &   $2^+_{\beta}$
& 4.1   &  5.4(12) \\
             &  1585              &            &$3^-_2$ &   $2^+_{gs}$
& 100    & 26.6(25) \\
             &                    &            &        &   $4^+_{gs}$
&  1.6    &  100(3)  \\
             &  1774              &            &$5^-_2$ &   $4^+_{gs}$
&  100    & 40(5)    \\
             &                    &            &        &   $6^+_{gs}$
&  2     &  100(4)  \\
\hline
\end{tabular}
\end{center}
\caption{Comparisons of relative intensities}
\label{t5}
\end{table}

\begin{table}
\begin{center}
\begin{tabular}{c l l l l l l }
\multicolumn{7}{l}{table \protect\ref{t5} continued.}\\ \hline
  Nucleus    &  $E_{level}(keV)$  & $K^{\pi}$  & $I_i$  &   $ I_f$ 
&   Cal.  &   Exp.   \\ \hline
  $^{156}$Gd &  1242              &  $1^-$     &$1^-_1$ &   $0^+_{gs}$
&   100   &  97.1(5) \\
             &                    &            &        &   $2^+_{gs}$
&  54     &  100(10) \\
             &                    &            &        &   $0^+_{\beta}$
&  7.0$\times10^{-6}$     &  0.07(2) \\
             &  1276              &            &$3^-_1$ &   $2^+_{gs}$
&  100    &  100(2)  \\
             &                    &            &        &   $4^+_{gs}$
&   87    &  44(2)   \\
             &  1320              &            &$2^-_1$ &   $2^+_{gs}$
&  100    &  100(3)  \\
             &                    &            &        &   $2^+_{\beta}$
& 7.0$\times10^{-5}$      &  0.21(2) \\
             &  1366              &  $0^-  $   &$1^-_2$ &   $0^+_{gs}$
&  61.1   &  54.5(3) \\
             &                    &            &        &   $2^+_{gs}$
&  100    &  100(1)  \\
             &                    &            &        &   $2^+_{\beta}$
&  0.09   &  0.08(3) \\
             &  1539              &            &$3^-_2$ &   $2^+_{gs}$
&  100    &  94(6)   \\
             &                    &            &        &   $4^+_{gs}$
&  86     &  100(6)  \\
             &  1638              &   $1^-$    &$7^-_1$ &   $6^+_{gs}$
&  100    & 100(3)   \\
             &                    &            &        &   $8^+_{gs}$
&  94     & $\approx 33$ \\
             &  1958              &            &$9^-_1$ &   $8^+_{gs}$
&  100    & 100(3)   \\
             &                    &            &        &   $10^+_{gs}$
&  82     & 7(3)     \\

  $^{158}$Gd &  977               &  $1^-$     &$1^-_1$ &   $0^+_{gs}$
&   100   &  100(5)  \\
             &                    &            &        &   $2^+_{gs}$
&  51     &  76(4)   \\
             &  1042              &            &$3^-_1$ &   $2^+_{gs}$
&  100    &  100(20) \\
             &                    &            &        &   $4^+_{gs}$
&   76    &  47.2(9) \\
             &  1176              &            &$5^-_1$ &   $4^+_{gs}$
&  100    &  100(6)  \\
             &                    &            &        &   $6^+_{gs}$
& 75.9    &  26.7(16)\\
             &  1260              &         &$2^+_{\beta}$ &$1^-_1$
&  5.1    &  5.1(46) \\
             &                    &            &        &   $2^-_1$
&  2.75   &  0.56(4) \\
             &                    &            &        &   $3^-_1$
&  2.01   &  6.9(6)  \\
             &  1263              &  $0^-  $   &$1^-_2$ &   $0^+_{gs}$
&  64     &  68(4)   \\
             &                    &            &        &   $2^+_{gs}$
&  100    &  100(6)  \\
             &  1403              &            &$3^-_2$ &   $2^+_{gs}$
&  100    &  100(6)  \\
             &                    &            &        &   $4^+_{gs}$
&  85     &  85(5)   \\
             &                    &            &        &   $2^+_{\beta}$
& 0.002   &  0.04(1) \\
             &  1407              &  $1^-$   &$4^+_{\beta}$ &$3^-_1$
&  19.9   &  19.9(12)\\
             &                    &            &        &   $4^-_1$
&  1.29   &  0.33(2) \\
             &                    &            &        &   $5^-_1$
&  5.21   &  6.6(5)  \\
             &  1639              &  $0^-  $   &$5^-_2$ &   $4^+_{gs}$
&  100    & 100(8)   \\
             &                    &            &        &   $6^+_{gs}$
&  61     & 33(5)     \\

  $^{160}$Gd &  1224              &  $0^-$     &$1^-_1$ &   $0^+_{gs}$
&   60.5  &  65.2(15)\\
             &                    &            &        &   $2^+_{gs}$
&  100    & 100.0(22)\\
             &  1290              &            &$3^-_1$ &   $2^+_{gs}$
&  100    &  100(2)  \\
             &                    &            &        &   $4^+_{gs}$
&   84    &  52(2)   \\
             &  1428              &            &$5^-_1$ &   $4^+_{gs}$
&  100    &  100(6)  \\
             &                    &            &        &   $6^+_{gs}$
&  55     &   $<45$     \\
             &  1640              &            &$7^-_1$ &   $6^+_{gs}$
&  100    &  28(11)  \\
             &                    &            &        &   $8^+_{gs}$
&  36     &  100(22) \\

  $^{160}$Dy &  1359              &  $1^-$     &$2^-_2$ &   $2^+_{gs}$
&   100   &  100(3)  \\
             &                    &            &        &   $2^+_{\gamma}$
&  1.00   & 17.95(12)\\
             &                    &            &        &   $3^+_{\gamma}$
&  0.88   & 11.59(5) \\
             &  1399              &            &$3^-_2$ &   $2^+_{gs}$
&  100    &  100.0(13) \\
             &                    &            &        &   $4^+_{gs}$
&   83.8  &  54.7(5)   \\
             &                    &            &        &   $2^+_{\gamma}$
&  0.43   & 0.81(3)  \\
             &                    &            &        &   $3^+_{\gamma}$
&  1.77   & 0.50(3)  \\
             &                    &            &        &   $4^+_{\gamma}$
&  0.65   & 0.26(3)  \\
             &  1535              &            &$4^-_2$ &   $4^+_{gs}$
&  100    &  100(9)  \\
             &                    &            &        &   $3^+_{\gamma}$
&  0.55   & 79.8(14)  \\
             &                    &            &        &   $4^+_{\gamma}$
&  1.22   & 13.4(6)  \\
             &                    &            &        &   $5^+_{\gamma}$
&  0.29   & 19.6(9)  \\
             &  1489              &  $^{*}0^-  $   &$1^-_2$ &   $0^+_{gs}$
&  60     &  69(14)   \\
             &                    &            &        &   $2^+_{gs}$
&  100    &  100    \\

  $^{162}$Dy &  1276              &  $0^-$     &$1^-_1$ &   $0^+_{gs}$
&   53    &  52(14)  \\
             &                    &            &        &   $2^+_{gs}$
&  100    & 100(4)   \\
             &  1358              &            &$3^-_1$ &   $2^+_{gs}$
&  84    &  100(30) \\
             &                    &            &        &   $4^+_{gs}$
&   100    &  46(4)   \\
             &  1519              &            &$5^-_1$ &   $4^+_{gs}$
&  100    & 100(20)   \\
             &                    &            &        &   $6^+_{gs}$
&  96     & 40(8)     \\
             &  1692              &   $1^-$    &$2^-_1$ &   $2^+_{gs}$
&  58    &  100(3)  \\
             &                    &            &        &   $3^+_{\gamma}$
&  100    &  49(6) \\
\hline
\end{tabular}
\end{center}
\end{table}

\end{document}